\documentclass[10pt,letterpaper]{article}
\usepackage{opex3}
\usepackage{color}
\usepackage{threeparttable}
\usepackage{amsmath}
\newcommand{\neff}{n_{\text{eff}}}

\begin{document}

\title{Tapered Simplified Modal Method for Analysis of Non-rectangular Gratings }

\author{Shuai Li$^{1,*}$, Changhe Zhou$^{3}$, and George Barbastathis$^{1,2}$}

\address{$^1$Department of Mechanical Engineering, Massachusetts Institute of Technology,\\ 77 Massachusetts Avenue, Cambridge, MA, 02139, USA\\
$^2$Singapore-MIT Alliance for Research and Technology (SMART) Centre,\\ 1 Create Way, 138602, Singapore\\
$^3$Shanghai Institute of Optics and Fine Mechanics, Chinese Academy of Science, \\ Shanghai, 201800, China}

\email{$^*$shuaili@mit.edu} %% email address is required



\begin{abstract}
The Simplified Modal Method (SMM) \cite{tishchenko2005phenomenological} provides a quick and intuitive way to analyze the performance of gratings of rectangular shapes. For non-rectangular shapes, a version of SMM has been developed \cite{zheng2008polarizing}, but it applies only to the Littrow-mounting incidence case and it neglects reflection. Here, we use the theory of mode-coupling in a tapered waveguide to improve SMM so that it applies to non-rectangular gratings at arbitrary angles of incidence. Moreover, this new $\lq\lq$Tapered Simplified Modal Method" (TSMM) allows us to properly account for reflected light. We present here the analytical development of the theory and numerical simulations, demonstrating the validity of the method.
\end{abstract}

 % REPLACE WITH CORRECT OCIS CODES FOR YOUR ARTICLE, MINIMUM OF TWO; Avoid using the OCIS codes for “General” or “General science” whenever possible.
%For a complete list of OCIS codes, visit: http://www.opticsinfobase.org/submit/ocis/

%%%%%%%%%%%%%%%%%%%%%%% References %%%%%%%%%%%%%%%%%%%%%%%%%
\bibliographystyle{osajnl}
\bibliography{reference}

\begin{thebibliography}{10}
\newcommand{\enquote}[1]{``#1''}

\bibitem{tishchenko2005phenomenological}
A.~Tishchenko, \enquote{Phenomenological representation of deep and high
  contrast lamellar gratings by means of the modal method,} Optical and quantum
  electronics \textbf{37}, 309--330 (2005).

\bibitem{zheng2008polarizing}
J.~Zheng, C.~Zhou, J.~Feng, and B.~Wang, \enquote{Polarizing beam splitter of
  deep-etched triangular-groove fused-silica gratings,} Optics letters
  \textbf{33}, 1554--1556 (2008).

\bibitem{park2012nanotextured}
K.-C. Park, H.~J. Choi, C.-H. Chang, R.~E. Cohen, G.~H. McKinley, and
  G.~Barbastathis, \enquote{Nanotextured silica surfaces with robust
  superhydrophobicity and omnidirectional broadband supertransmissivity,} ACS
  nano \textbf{6}, 3789--3799 (2012).

\bibitem{karagodsky2010theoretical}
V.~Karagodsky, F.~G. Sedgwick, and C.~J. Chang-Hasnain, \enquote{Theoretical
  analysis of subwavelength high contrast grating reflectors,} Optics express
  \textbf{18}, 16973--16988 (2010).

\bibitem{karagodsky2012physics}
V.~Karagodsky and C.~J. Chang-Hasnain, \enquote{Physics of near-wavelength high
  contrast gratings,} Optics express \textbf{20}, 10888--10895 (2012).

\bibitem{feng2008modal}
J.~Feng, C.~Zhou, J.~Zheng, and B.~Wang, \enquote{Modal analysis of deep-etched
  low-contrast two-port beam splitter grating,} Optics Communications
  \textbf{281}, 5298--5301 (2008).

\bibitem{clausnitzer2005intelligible}
T.~Clausnitzer, T.~K{\"a}mpfe, E.-B. Kley, A.~T{\"u}nnermann, U.~Peschel,
  A.~Tishchenko, and O.~Parriaux, \enquote{An intelligible explanation of
  highly-efficient diffraction in deep dielectric rectangular transmission
  gratings,} Optics Express \textbf{13}, 10448--10456 (2005).

\bibitem{garcia2003finite}
S.~G. Garc{\i}a, A.~R. Bretones, B.~G. Olmedo, and R.~G. Mart{\i}n,
  \enquote{Finite difference time domain methods,} Time Domain Techniques in
  Computational Electromagnetics  (2003).

\bibitem{moharam1982diffraction}
M.~Moharam and T.~K. Gaylord, \enquote{Diffraction analysis of dielectric
  surface-relief gratings,} JOSA \textbf{72}, 1385--1392 (1982).

\bibitem{botten1981dielectric}
I.~Botten, M.~Craig, R.~McPhedran, J.~Adams, and J.~Andrewartha, \enquote{The
  dielectric lamellar diffraction grating,} Journal of Modern Optics
  \textbf{28}, 413--428 (1981).

\bibitem{li1993multilayer}
L.~Li, \enquote{Multilayer modal method for diffraction gratings of arbitrary
  profile, depth, and permittivity,} JOSA A \textbf{10}, 2581--2591 (1993).

\bibitem{li1996formulation}
L.~Li, \enquote{Formulation and comparison of two recursive matrix algorithms
  for modeling layered diffraction gratings,} JOSA A \textbf{13}, 1024--1035
  (1996).

\bibitem{cotter1995scattering}
N.~Cotter, T.~Preist, and J.~Sambles, \enquote{Scattering-matrix approach to
  multilayer diffraction,} JOSA A \textbf{12}, 1097--1103 (1995).

\bibitem{snyder1970coupling}
A.~W. Snyder, \enquote{Coupling of modes on a tapered dielectric cylinder,}
  Microwave Theory and Techniques, IEEE Transactions on \textbf{18}, 383--392
  (1970).

\bibitem{sheng1982exact}
P.~Sheng, R.~Stepleman, and P.~Sanda, \enquote{Exact eigenfunctions for
  square-wave gratings: Application to diffraction and surface-plasmon
  calculations,} Physical Review B \textbf{26}, 2907 (1982).

\bibitem{marcuvitz1951representation}
N.~Marcuvitz and J.~Schwinger, \enquote{On the representation of the electric
  and magnetic fields produced by currents and discontinuities in wave guides.
  i,} Journal of Applied Physics \textbf{22}, 806--819 (1951).

\bibitem{jing2013enhancement}
X.~Jing, S.~Jin, J.~Zhang, Y.~Tian, P.~Liang, H.~Shu, L.~Wang, and Q.~Dong,
  \enquote{Enhancement of the accuracy of the simplified modal method for
  designing a subwavelength triangular grooves grating,} Optics letters
  \textbf{38}, 10--12 (2013).

\bibitem{sun2015unified}
Z.~Sun, C.~Zhou, H.~Cao, and J.~Wu, \enquote{Unified beam splitter of fused
  silica grating under the second bragg incidence,} JOSA A \textbf{32},
  1952--1957 (2015).

\bibitem{johnson2002adiabatic}
S.~G. Johnson, P.~Bienstman, M.~Skorobogatiy, M.~Ibanescu, E.~Lidorikis, and
  J.~Joannopoulos, \enquote{Adiabatic theorem and continuous coupled-mode
  theory for efficient taper transitions in photonic crystals,} Physical review
  E \textbf{66}, 066608 (2002).

\bibitem{yang2015evaluation}
F.~Yang and Y.~Li, \enquote{Evaluation and improvement of simplified modal
  method for designing dielectric gratings,} Optics express \textbf{23},
  31342--31356 (2015).

\end{thebibliography}

%%%%%%%%%%%%%%%%%%%%%%%%%%  body  %%%%%%%%%%%%%%%%%%%%%%%%%%
\section{Introduction}
Gratings have become increasingly important in a large number of applications such as omnidirectional broadband transmitters \cite{park2012nanotextured}, high-contrast broadband reflectors \cite{karagodsky2010theoretical, karagodsky2012physics}, two-port beam splitters \cite{feng2008modal} and polarizing beam splitters \cite{zheng2008polarizing}. To facilitate the design process of those optical structures, an efficient and accurate computation method is of great importance. Within the past decades, several methods have been proposed to compute the optical field inside and outside the nanostructures \cite{clausnitzer2005intelligible}. Among them, the Finite Difference Time Domain method (FDTD) \cite{garcia2003finite} and Rigorous Coupled Wave Analysis (RCWA) \cite{moharam1982diffraction} are the widest applied nowadays. FDTD and RCWA numerically solve the Maxwell's equations in the time domain and frequency domain, respectively, and have shown high accuracy in practice. However, both methods give little insight into the physical process that takes place inside the grating region. Also, FDTD and RCWA can not be easily inverted to solve the inverse diffraction problem.

The modal method \cite{botten1981dielectric} is a possible alternative to FDTD and RCWA. Unlike these two methods, the modal method not only solves the Maxwell's equations, but also shows us an intuitive picture about the physical process inside the grating region. In addition, it can be easily inverted to solve the inverse diffraction problem. For gratings with small period, only a few (usually $2\sim3$) propagating modes inside the grating region exist and the diffraction process is dominated by the interference between them. In this way, a Simplified Modal Method (SMM) \cite{clausnitzer2005intelligible} was proposed, which further simplifies the computation. Moreover, by taking advantage of the two-beam interference mechanism under Littrow-mounting \cite{clausnitzer2005intelligible} and the effective refractive index matching \cite{tishchenko2005phenomenological} at the input and output interface, we can quickly predict the optical properties of the grating using SMM. In this way, SMM is well suited to the design of gratings. 

The modal method was originally derived based on rectangular gratings \cite{botten1981dielectric}. In order to extend the approach to gratings of arbitrary profile, a multilayer modal method (MMM) \cite{li1993multilayer} was proposed. However, in MMM the mode propagation between layers was obtained using a R-matrix algorithm which does not have an intuitive physical interpretation \cite{li1996formulation}. As a result, the most advantageous aspect of the modal method over FDTD and RCWA was compromised. The S-matrix algorithm \cite{cotter1995scattering} was another approach to deal with light propagation in a multilayer structures. It has a clear physical meaning but is not compatible with the modal method as the matrix would become singular in some conditions \cite{li1996formulation}. For SMM, one modified version \cite{zheng2008polarizing} has been developed for the analysis of non-rectangular gratings with the help of the two-beam interference mechanism, but it only works under the Littrow-mounting illumination condition. Moreover, reflections at both the input and output interface are neglected in that case. Above all, no modal method can be applied to non-rectangular gratings at arbitrary angles of incidence without losing its clear physical meaning.  Therefore, the application of the modal method in practice is limited nowadays as most optical nanostructures are of non-rectangular shapes.

In this paper, we propose the Tapered Simplified Modal Method (TSMM) as the first physically intuitive formulation of SMM that can be applied to non-rectangular gratings at arbitrary illumination angles. In TSMM, we first discretize the non-rectangular grating along the axial direction and then apply  conventional SMM theory to each discretized layer. The relationship between each layer, unlike \cite{li1993multilayer} where the R-matrix was used, is found using tapered mode-coupling analysis. 

\section{Main Theory}
We begin by discretizing the non-rectangular grating along the axial direction, as shown in Fig. 1. This procedure is the same as that in the MMM \cite{li1993multilayer}, and it is applicable to gratings of arbitrary shapes and at arbitrary illumination angles. Here, for the sake of simplicity, we illustrate the theory using a triangular grating. We assume $n_{1}$, $n_{2}$, $n_{b}$ and $n_{g}$ are the refractive indices of the input medium, the substrate, the ridge of the grating and the groove of the grating, respectively, $\varphi_{in}$ is the incident angle of the illumination light, and $d$ and $H$ are the period and groove depth of the grating, respectively.

\begin{figure}[h]
\centering\includegraphics[width=7cm]{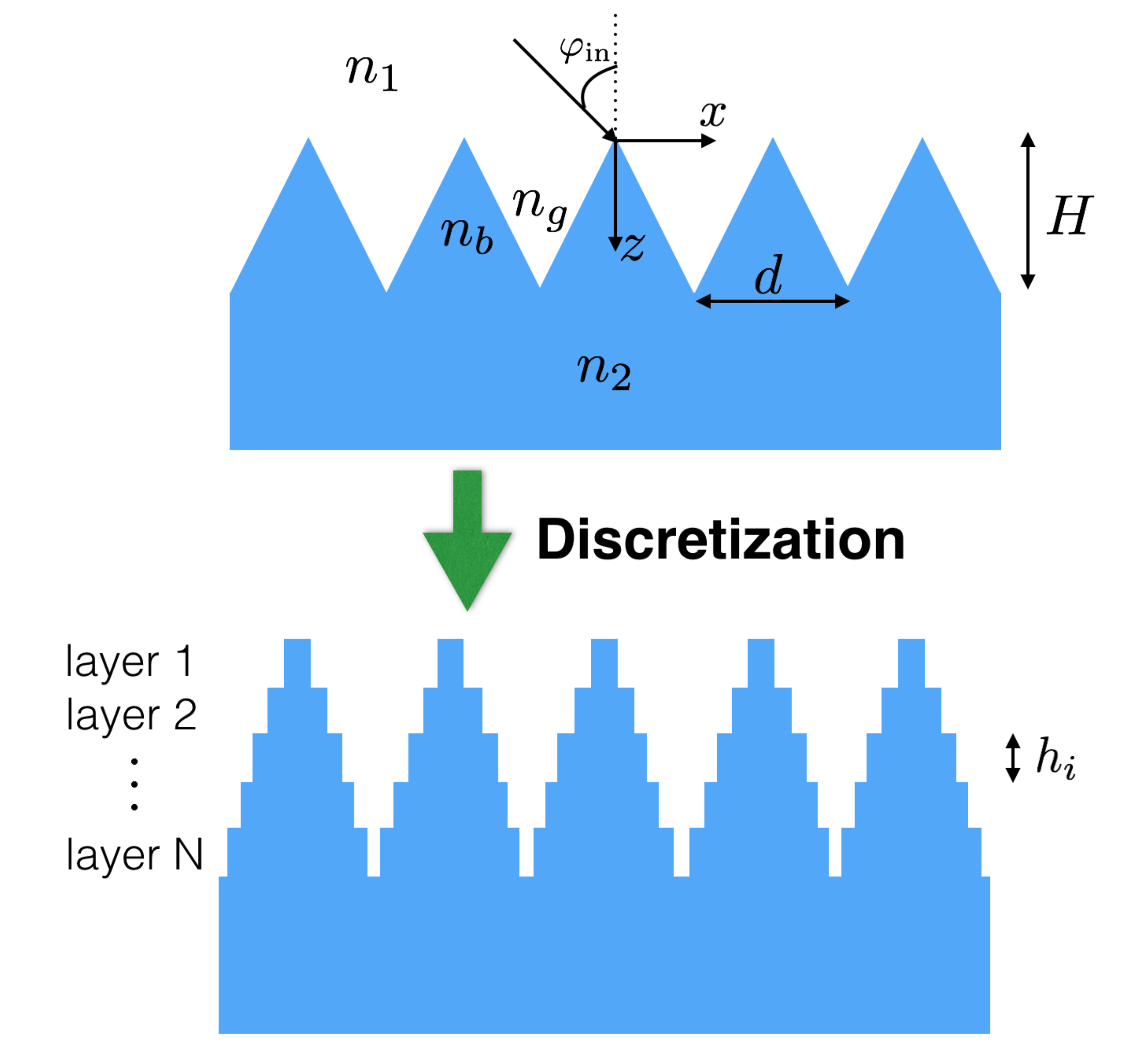}
\caption{Discretization of non-rectangular grating}
\end{figure} 

The non-rectangular grating is discretized into $N$ layers, each of groove depth $h_{i}=H/N$. These elemental gratings can now be considered as rectangular gratings. According to the modal theory, for a rectangular grating, the transverse field inside the grating region can be represented as the sum of grating modes propagating up and down the grating \cite{tishchenko2005phenomenological}:

\begin{equation}
\begin{cases}
&E_{t}(x,z)=\sum_{q=1}^{M}[a_{q}\text{exp}(\Gamma_{q}z)e_{q}(x)+a_{-q}\text{exp}(\Gamma_{-q}z)e_{-q}(x)],\\
&\\
&H_{t}(x,z)=\sum_{q=1}^{M}[a_{q}\text{exp}(\Gamma_{q}z)h_{q}(x)+a_{-q}\text{exp}(\Gamma_{-q}z)h_{-q}(x)].\\
\end{cases}
\label{eq:refname1}
\end{equation}

Here,  $M$ is the number of modes that we use to approximate the grating field; $a_{q}$ and $a_{-q}$ are the amplitude coefficients for the $q$-th forward mode and $q$-th backward mode, respectively; and $\Gamma_{q}$ and $\Gamma_{-q}$ are the modal propagation constants for the $q$-th forward mode and $q$-th backward mode, respectively. In SMM, we only consider the propagating mode, {\it i.e.} $\Gamma_{q}$ and $\Gamma_{-q}$ are purely imaginary. The modal propagation constants can be computed as: $\Gamma_{q}=ik_{0}\neff(q)$ and $\Gamma_{-q}=-\Gamma_{q}$. Here, $k_{0}=2\pi/\lambda$ and $\neff(q)$ is the effective refractive index for the $q$-th mode, which is the $q$-th largest possible value of $\neff$ that satisfies the transcendent equation \cite {botten1981dielectric}:

\begin{equation}
\cos{(\alpha d)}=\cos{(\beta fd)}\cdot\cos{[\gamma (1-f)d]}-\frac{\beta^2+\tau^2\gamma^2}{2\beta\gamma\tau}\sin{(\beta fd)}\sin{[\gamma(1-f)d]},
\label{eq:refname1}
\end{equation}
where $\alpha=k_{0}n_{1}\sin{\varphi_{in}}$,  $\beta=k_{0}\sqrt{n_{b}^{2}-\neff^2}$, $\gamma=k_{0}\sqrt{n_{g}^{2}-\neff^2}$.

Eq. (2) is obtained through the field continuity condition at the boundary between ridges and grooves. Therefore, this equation should be polarization dependent and the polarization factor $\tau$ is defined as:

\begin{equation}
\tau=
\begin{cases}
\frac{n_{b}^2}{n_{g}^2}, &\text{For TM polarization} (H_{x}=H_{z}=0);\\
1, &\text{For TE polarization} (E_{x}=E_{z}=0);\\
\end{cases}
\label{eq:refname1}
\end{equation}

In eq. (1), $e_{q}(x), e_{-q}(x), h_{q}(x), h_{-q}(x)$ are the transverse parts of the electric and magnetic field function for the $q$-th forward mode and $q$-th backward mode, respectively. They follow the symmetry relationship \cite {snyder1970coupling}: $e_q(x)=e_{-q}(x)$, $h_q(x)=-h_{-q}(x)$. Then, eq. (1) can be rewritten as:

\begin{equation}
\begin{cases}
&E_{t}(x,z)=\sum_{q=1}^{M}[A_{q}(z)+A_{-q}(z)]\cdot e_q(x),\\
&\\
&H_{t}(x,z)=\sum_{q=1}^{M}[A_{q}(z)-A_{-q}(z)]\cdot h_q(x),\\
\end{cases}
\label{eq:refname1}
\end{equation}
where $A_{q}(z)$ is the modal amplitude coefficient:

\begin{equation}
A_{q}(z)=a_{q}e^{\Gamma_{q}z}.
\label{eq:refname1}
\end{equation}

The distributions of $e_{q}(x)$ and $h_{q}(x)$ are determined through the field continuity condition at the boundary between ridges and grooves. Hence, the mode functions are also polarization dependent. Specifically \cite{sheng1982exact},

\begin{itemize}
\item{for TE polarization
\begin{equation}
e_{q}(x)=
\begin{cases}
\cos{[\beta(x+\frac{fd}{2})]}+iV_{0}\frac{\tau k_{0}}{\beta}\sin{[\beta(x+\frac{fd}{2})]}, \quad &\text{For} |x|\leq\frac{fd}{2};\\
U_{1}\cos{[\gamma(x-\frac{fd}{2})]}+iV_{1}\frac{k_{0}}{\gamma}\sin{[\gamma(x-\frac{fd}{2})]}, \quad &\text{For} \frac{fd}{2}\leq |x|\leq(1-\frac{f}{2})d;\\
\end{cases}
\label{eq:refname1}
\end{equation}

\begin{equation}
h_{q}(x)=\frac{k_{0}\cdot \neff(q)}{\omega\mu}e_{q}(x);
\end{equation}}

\item{for TM polarization
\begin{equation}
e_{q}(x)=\frac{-k_{0}\cdot \neff(q)}{\omega\epsilon}h_{q}(x);
\end{equation}

\begin{equation}
h_{q}(x)=
\begin{cases}
\cos{[\beta(x+\frac{fd}{2})]}+iV_{0}\frac{\tau k_{0}}{\beta}\sin{[\beta(x+\frac{fd}{2})]}, \quad &\text{For} |x|\leq\frac{fd}{2};\\
U_{1}\cos{[\gamma(x-\frac{fd}{2})]}+iV_{1}\frac{k_{0}}{\gamma}\sin{[\gamma(x-\frac{fd}{2})]}, \quad &\text{For} \frac{fd}{2}\leq |x|\leq(1-\frac{f}{2})d;\\
\end{cases}
\label{eq:refname1}
\end{equation}}
\end{itemize}
Here, 
\begin{equation}
\begin{aligned}
&V_{0}=[exp(ik_{0}d\sin{\varphi_{in}})-M]/P,\\
&V_{1}=\frac{i}{k_{0}}\frac{\beta}{\tau}\sin{(\beta fd)}+V_{0}\cos{(\beta fd)},\\
&U_{1}=\cos{(\beta fd)}+iV_{0}\frac{\tau k_{0}}{\beta}\sin{\beta fd},\\
&M=\cos{(\beta fd)}\cos{[\gamma(1-f)d]}-\frac{\beta}{\tau\gamma}\sin{[\gamma(1-f)d]}\sin{(\beta fd)},\\
&P=ik_{0}\{\frac{1}{\gamma}\cos{(\beta fd)}\sin{[\gamma(1-f)d]}+\frac{\tau}{\beta}\cos{[\gamma(1-f)d]}\sin{(\beta fd)}\}.\\
\end{aligned}
\label{eq:refname1}
\end{equation}
The above expressions for $e_{q}(x)$ and $h_{q}(x)$ are for one period and they can readily be extended to the whole domain through  Bloch's Theorem \cite {clausnitzer2005intelligible}.

As illustrated in Fig. 1, each layer of the discretized grating will have the same period, but different duty cycle. Therefore, according to eq. (2), each layer will have different effective refractive indices $\neff$ for the same mode. Moreover, when $f\to 0$, $\neff(1)\to n_{g}\cos{\varphi_{in}}$;  $f\to 1$, $\neff(1)\to n_{b}\cos{\varphi_{d}}$, where $\varphi_{d}$ is the diffraction angle inside the substrate. Hence, the discretized grating can also be considered as a progression of effective refractive indices along the axial direction, as shown in Fig. 2.

\begin{figure}[h]
\centering\includegraphics[width=7cm]{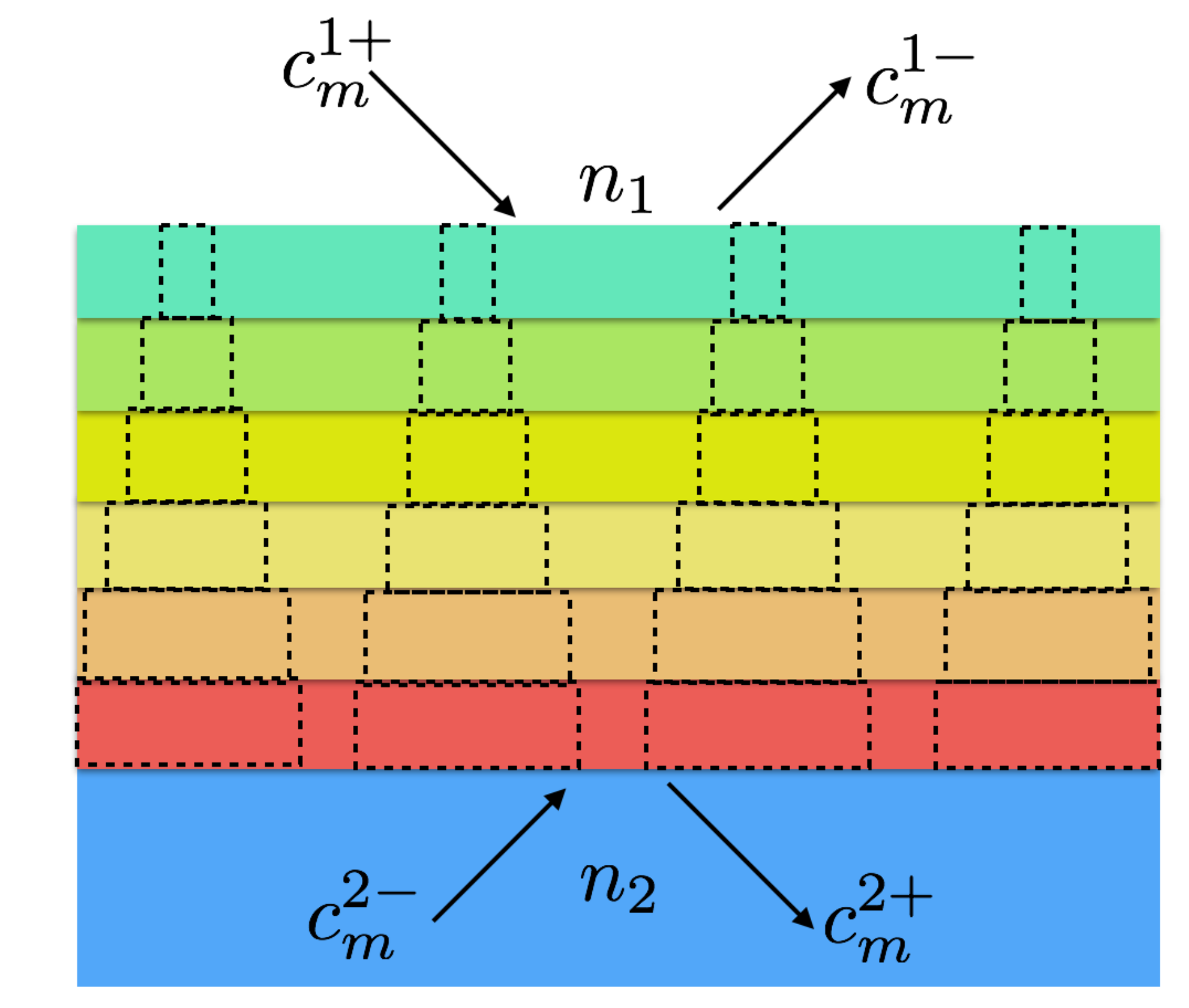}
\caption{Scheme for estimating the discretized grating along the axial direction as a progression of effective refractive indices }
\end{figure} 

After discretization, the field inside each layer can be expressed as:
\begin{equation}
\begin{cases}
&E_{t}^{j}(x)=\sum_{q=1}^{M}[A_{q}^{j}+A_{-q}^{j}]\cdot e_q^{j}(x),\\
&H_{t}^{j}(x)=\sum_{q=1}^{M}[A_{q}^{j}-A_{-q}^{j}]\cdot h_q^{j}(x),\\
\end{cases}
\quad j=1,2,\cdots,N.
\label{eq:refname1}
\end{equation}
Here, the superscript $j$ denotes the $j$-th layer.

Using the field continuity condition for $E_{t}$ and $H_{t}$ at the input interface ($z=0$) and the output interface ($z=H$), we can construct the following four boundary conditions. Here, for simplicity, we only illustrate the TE polarization case (the TM case is derived similarly).

\begin{itemize}
\item{
At the input interface ($j=1$):
\begin{equation}
\sum_{m=-\infty}^{\infty}(c_{m}^{1+}+c_{m}^{1-})exp(ik_{xm}x)=\sum_{q=1}^{M}(A_{q}^{1}+A_{-q}^{1})\cdot e_q^{1}(x);
\label{eq:refname1}
\end{equation}

\begin{equation}
\sum_{m=-\infty}^{\infty}\frac{k_{m}^{1}}{\omega\mu}(c_{m}^{1+}-c_{m}^{1-})exp(ik_{xm}x)=\sum_{q=1}^{M}(A_{q}^{1}-A_{-q}^{1})\cdot h_q^{1}(x);
\label{eq:refname1}
\end{equation}}

\item{
At the output interface ($j=N$):

\begin{equation}
\sum_{m=-\infty}^{\infty}(c_{m}^{2+}+c_{m}^{2-})exp(ik_{xm}x)=\sum_{q=1}^{M}(A_{q}^{N}+A_{-q}^{N})\cdot e_q^{N}(x);
\label{eq:refname1}
\end{equation}

\begin{equation}
\sum_{m=-\infty}^{\infty}\frac{k_{m}^{2}}{\omega\mu}(c_{m}^{2+}-c_{m}^{2-})exp(ik_{xm}x)=\sum_{q=1}^{M}(A_{q}^{N}-A_{-q}^{N})\cdot h_q^{N}(x);
\label{eq:refname1}
\end{equation}}
\end{itemize}
Here, $k_{xm}=\frac{2\pi n_{1}\sin{\varphi_{in}}}{\lambda}+\frac{m\cdot2\pi}{d}$ is the transverse wavenumber of the $m$-th order diffraction beam; $k_{m}^{1}=\sqrt{\left(\frac{2\pi n_{1}}{\lambda}\right)^2-(k_{xm})^2}$ and $k_{m}^{2}=\sqrt{\left(\frac{2\pi n_{2}}{\lambda}\right)^2-(k_{xm})^2}$ are the axial wavenumbers of the $m$-th order diffraction beam inside the input medium and the substrate, respectively; and $c_{m}^{1+}$ and $c_{m}^{1-}$ are the amplitude coefficients of the $m$-th order diffraction beam propagating forwards and backwards inside the input medium, respectively. Similarly, $c_{m}^{2+}$ and $c_{m}^{2-}$ are the amplitude coefficients of the $m$-th order diffraction beam propagating forwards and backwards inside the substrate, respectively. By considering the backward propagating beam, our method accounts for reflection, which is neglected in conventional SMM.

Combining eq. (12) $\sim$(15), we readily obtain:

\begin{equation}
c_{n}^{1+}=\frac{\omega\mu}{2k_{n}^{1}}\left\{\sum_{q=1}^{M}\left(\frac{k_{n}^{1}}{\omega\mu}F_{q}^{n(1)}+G_{q}^{n(1)}\right)A_{q}^{1}+\sum_{q=1}^{M}\left(\frac{k_{n}^{1}}{\omega\mu}F_{q}^{n(1)}-G_{q}^{n(1)}\right)A_{-q}^{1}\right\},
\label{eq:refname1}
\end{equation}

\begin{equation}
c_{n}^{1-}=\frac{\omega\mu}{2k_{n}^{1}}\left\{\sum_{q=1}^{M}\left(\frac{k_{n}^{1}}{\omega\mu}F_{q}^{n(1)}-G_{q}^{n(1)}\right)A_{q}^{1}+\sum_{q=1}^{M}\left(\frac{k_{n}^{1}}{\omega\mu}F_{q}^{n(1)}+G_{q}^{n(1)}\right)A_{-q}^{1}\right\},
\label{eq:refname1}
\end{equation}

\begin{equation}
c_{n}^{2+}=\frac{\omega\mu}{2k_{n}^{2}}\left\{\sum_{q=1}^{M}\left(\frac{k_{n}^{2}}{\omega\mu}F_{q}^{n(N)}+G_{q}^{n(N)}\right)A_{q}^{N}+\sum_{q=1}^{M}\left(\frac{k_{n}^{2}}{\omega\mu}F_{q}^{n(N)}-G_{q}^{n(N)}\right)A_{-q}^{N}\right\},
\label{eq:refname1}
\end{equation}

\begin{equation}
c_{n}^{2-}=\frac{\omega\mu}{2k_{n}^{2}}\left\{\sum_{q=1}^{M}\left(\frac{k_{n}^{2}}{\omega\mu}F_{q}^{n(N)}-G_{q}^{n(N)}\right)A_{q}^{N}+\sum_{q=1}^{M}\left(\frac{k_{n}^{2}}{\omega\mu}F_{q}^{n(N)}+G_{q}^{n(N)}\right)A_{-q}^{N}\right\}.
\label{eq:refname1}
\end{equation}
Here,

\begin{equation}
F_{q}^{n(j)}=\frac{1}{dc}\int_{0}^{dc}e_{q}^{j}(x)\cdot exp(-ik_{xn}x)dx, \quad j=1,N,
\end{equation}

\begin{equation}
G_{q}^{n(j)}=\frac{1}{dc}\int_{0}^{dc}h_{q}^{j}(x)\cdot exp(-ik_{xn}x)dx, \quad j=1,N.
\end{equation}
The value of the integral period $dc$ is the same as the period of the field function, which in turn depends on the incident angle:  $dc=\frac{N_{1}\cdot\lambda}{n_{1}\sin{\varphi_{in}}}=N_{2}\cdot d$. Here, $N_{1}$ is chosen to be the smallest nonnegative integer that makes $N_{2}$ a positive integer. Under normal incidence condition ($\varphi_{in}=0$), we obtain $dc=d$. Under Littrow-mounting illumination condition ($\sin{\varphi_{in}}=\lambda/2n_{1}d$), we obtain $dc=2d$.
 
 Looking at eq. (16) $\sim$ (19), we can find that if we can relate $A_{q}^{N}$ and $A_{-q}^{N}$ with $A_{q}^{1}$ and $A_{-q}^{1}$, then all the unknown amplitudes $c_{n}^{1-}, c_{n}^{2+}, A_{q}^{1}$ and $A_{-q}^{1}$ can be solved via incident amplitudes $c_{n}^{1+}$ and $c_{n}^{2-}$.  We now proceed to find the relationship between $A_{q}^{N}, A_{-q}^{N}$ and $A_{q}^{1}, A_{-q}^{1}$ by using tapered mode-coupling theory \cite{snyder1970coupling}. 
 
Let's start with Maxwell's equations:
\begin{equation}
\begin{aligned}
&\nabla\times H=i\omega\epsilon E,\\
&\nabla\times E=-i\omega\mu H.\\
\end{aligned}
\end{equation}

We write the electric and magnetic fields $E$ and $H$ as the superpositions of their respective transverse parts $E_{t}, H_{t}$ and longitudinal parts $E_{z}, H_{z}$:
 \begin{equation}
 \begin{aligned}
&E=(E_{t}+E_{z})e^{jwt},\\
&H=(H_{t}+H_{z})e^{jwt}.\\
\end{aligned}
\end{equation}

Then, Maxwell's equations (22) are recast into transmission line form \cite{marcuvitz1951representation}:
 \begin{equation}
 \begin{aligned}
&\frac{-\partial E_{t}}{\partial z}=i\omega\left(\mu I+\frac{1}{\omega^2}\nabla_{t}\frac{1}{\epsilon}\nabla_{t}\right)\cdot\left(H_{t}\times\hat{z}\right),\\
&\frac{-\partial H_{t}}{\partial z}=i\omega\left(\epsilon I+\frac{1}{\omega^2}\nabla_{t}\frac{1}{\mu}\nabla_{t}\right)\cdot\left(\hat{z}\times E_{t}\right),\\
\end{aligned}
\end{equation}
where $\nabla_{t}=\nabla-\hat{z}\frac{\partial}{\partial z}$.

Substituting eq.(4) into the above transmission line equation, we can obtain the following set of differential equations:
\begin{equation}
\frac{dA_{q}}{dz}-\Gamma_{q}A_{q}=\frac{1}{2}\sum_{p=-M}^{M}A_{p}(K_{pq}+\tilde{K}_{qp}).
\end{equation}
Here, $K_{pq}$ and $\hat{K}_{pq}$ are two types of coupling coefficients:
 \begin{equation}
 \begin{aligned}
&K_{pq}=\int_{0}^{dc}\hat{z}\cdot e_{p}(x)\times\frac{\partial h^{*}_{q}(x)}{\partial z}dx,\\
&\tilde{K}_{pq}=\int_{0}^{dc}\hat{z}\cdot \frac{\partial e^{*}_{p}(x)}{\partial z}\times h_{q}(x)dx.\\
\end{aligned}
\end{equation}

This set of differential equations is physically intuitive, since it exactly represents the mode coupling process inside the grating region. By solving it, we can obtain the amplitude coefficients for each mode at any axial position and can then relate $A_{q}^{N}$ and $A_{-q}^{N}$ with $A_{q}^{1}$ and $A_{-q}^{1}$ through a simple matrix multiplication:
\begin{equation}
\begin{bmatrix}
A_{q}^{N}\\
A_{-q}^{N}\\
\end{bmatrix}
=M_{t}\cdot
\begin{bmatrix}
A_{q}^{1}\\
A_{-q}^{1}\\
\end{bmatrix}.
\end{equation}
Here, $M_{t}$ is the transfer matrix.

Then, the boundary continuity equations eq. (16) $\sim$ (19) can be solved as:
\begin{itemize}
\item{input:
\begin{equation}
\begin{bmatrix}
c_{n}^{1+}\\
c_{n}^{2-}\\
\end{bmatrix}
=\sum_{q=1}^{M}
\begin{bmatrix}
\frac{F_{q}^{n(1)}}{2}+\frac{\omega\mu}{2k_n^1}G_{q}^{n(1)} &\frac{F_{q}^{n(1)}}{2}-\frac{\omega\mu}{2k_n^1}G_{q}^{n(1)}\\
U_{q} &V_{q}\\
\end{bmatrix}
\begin{bmatrix}
A_{q}^{1}\\
A_{-q}^{1}\\
\end{bmatrix};
\end{equation}

\begin{equation}
\sum_{q=1}^{M}[U_{q},V_{q}]=\sum_{q=1}^{M}\left[\frac{F_{q}^{n(N)}}{2}-\frac{\omega\mu}{2k_n^2}G_{q}^{n(N)}, \frac{F_{q}^{n(N)}}{2}+\frac{\omega\mu}{2k_n^2}G_{q}^{n(N)}\right]\cdot M_{t};
\end{equation}}

\item{output:
\begin{equation}
\begin{bmatrix}
c_{n}^{1-}\\
c_{n}^{2+}\\
\end{bmatrix}
=\sum_{q=1}^{M}
\begin{bmatrix}
\frac{F_{q}^{n(1)}}{2}-\frac{\omega\mu}{2k_n^1}G_{q}^{n(1)} &\frac{F_{q}^{n(1)}}{2}+\frac{\omega\mu}{2k_n^1}G_{q}^{n(1)}\\
U'_{q} &V'_{q}\\
\end{bmatrix}
\begin{bmatrix}
A_{q}^{1}\\
A_{-q}^{1}\\
\end{bmatrix};
\end{equation}

\begin{equation}
\sum_{q=1}^{M}[U'_{q},V'_{q}]=\sum_{q=1}^{M}\left[\frac{F_{q}^{n(N)}}{2}+\frac{\omega\mu}{2k_n^2}G_{q}^{n(N)}, \frac{F_{q}^{n(N)}}{2}-\frac{\omega\mu}{2k_n^2}G_{q}^{n(N)}\right]\cdot M_{t};
\end{equation}}
\end{itemize}

In this way, we obtain all the unknown amplitudes $c_{n}^{1-}, c_{n}^{2+}$ and thus the diffraction field and the reflection field of the grating.

\section{Simulations and Results}
Based on TSMM theory, we performed simulations on several different non-rectangular gratings under different illumination angles. We analyzed the simulation results and demonstrated the validity of the proposed approach numerically.
\subsection {Littrow-mounting illumination condition}
The first case we consider is the Littrow-mounting illumination case. This is the only case that conventional SMM can be applied for a non-rectangular grating. As an example, we analyze a triangular grating with $d=L=600nm, n_{1}=n_{g}=1,n_{2}=n_{b}=1.45$. Here, $L$ is the width at the bottom of the triangle. The illumination wavelength is set to be $\lambda=1064nm$. Hence, the illumination angle should be $\varphi_{in}=\sin^{-1}\left(\frac{\lambda}{2n_{1}d}\right)=62.46^{\circ}$ to fulfill the Littrow-mounting condition. We computed the diffraction efficiencies of the 0th order and -1st order diffraction beams with different values of aspect ratio $H/d$ of the triangular grating. The result is shown in Fig. 3.

The simulation result matches well with the RCWA result for the same structure provided in reference \cite{jing2013enhancement}. As shown in Fig. 3, the result indicates the interference mechanism \cite{clausnitzer2005intelligible} between the two propagating modes inside the triangular grating in the Littrow-mounting illumination condition (where 0th and $-1$st order diffraction beam are symmetric), which is very similar to a Mach-Zehnder Interferometer. The high transmission ratio in this case can be explained by the fact that when $f\to 0$, $\neff(1)\to n_{g}\cos{\varphi_{in}}$;  whereas when $f\to 1$, $\neff(1)\to n_{b}\cos{\varphi_{d}}$. Here, in this structure, the duty cycle at the top of the triangle goes to 0 and the the duty cycle at the bottom of the triangle goes to 1. Hence, at both the input interface and the output interface, the reflection caused by the refractive index mismatch is very low.
\begin{figure}[h]
\centering\includegraphics[width=9.5cm]{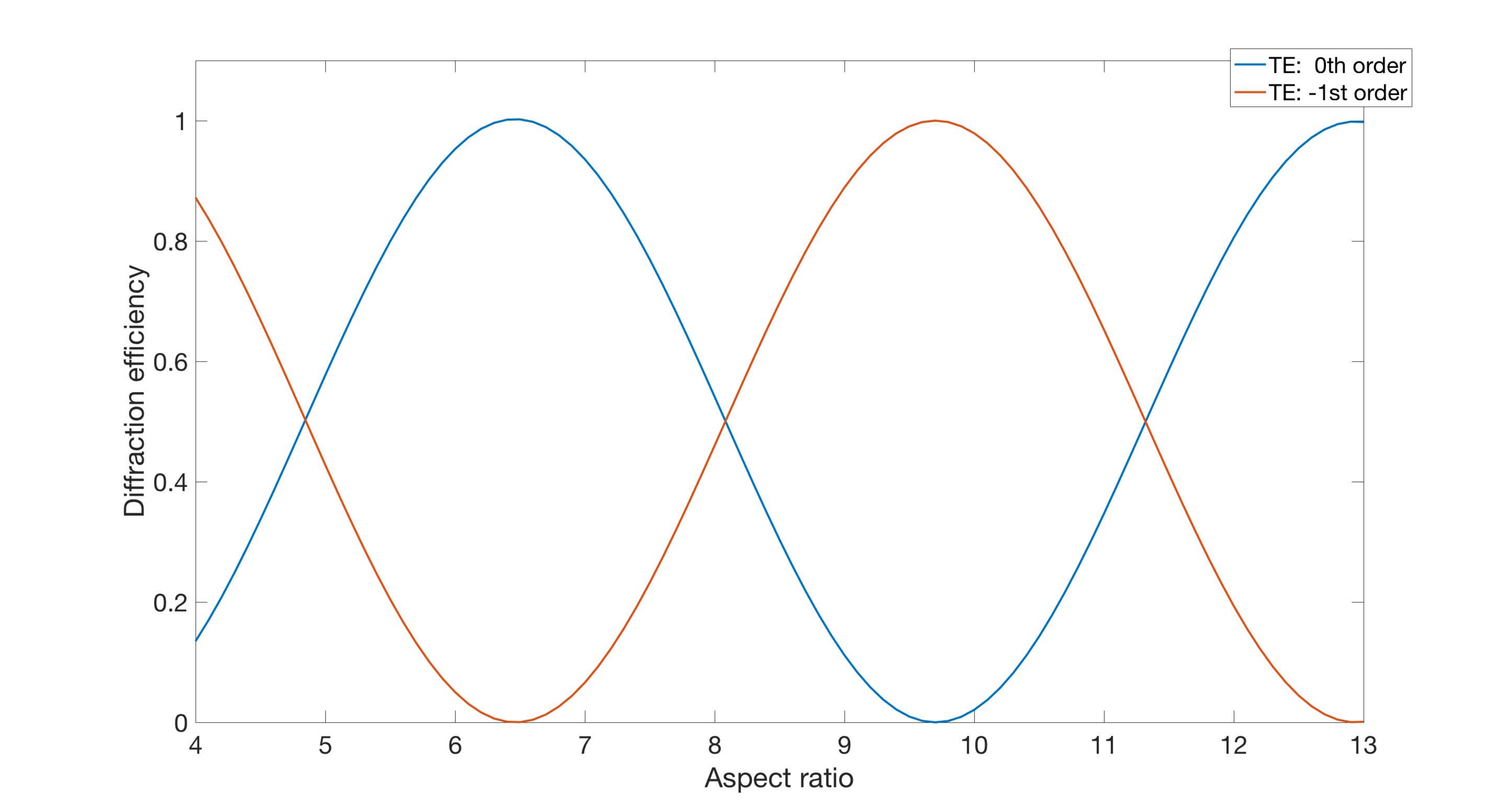}
\caption{Littrow-mounting illumination condition}
\end{figure}

\subsection {Non-Littrow-mounting illumination condition}
This is a case that cannot be handled by conventional SMM, but TSMM as presented here has no problems with. As example, we consider a triangular grating with 
$d=L=900nm, n_{1}=n_{g}=1,n_{2}=n_{b}=1.45$. The illumination wavelength is $\lambda=1200nm$. The illumination angle $\varphi_{in}=\sin^{-1}\left(\frac{\lambda}{3n_{1}d}\right)=26.39^{\circ}$, which is no longer the Littrow-mounting case. The diffraction efficiencies of the 0th order and -1st order diffraction beams with respect to varying values of the aspect ratio are shown in Fig. 4.
\begin{figure}[h]
\centering\includegraphics[width=9.5cm]{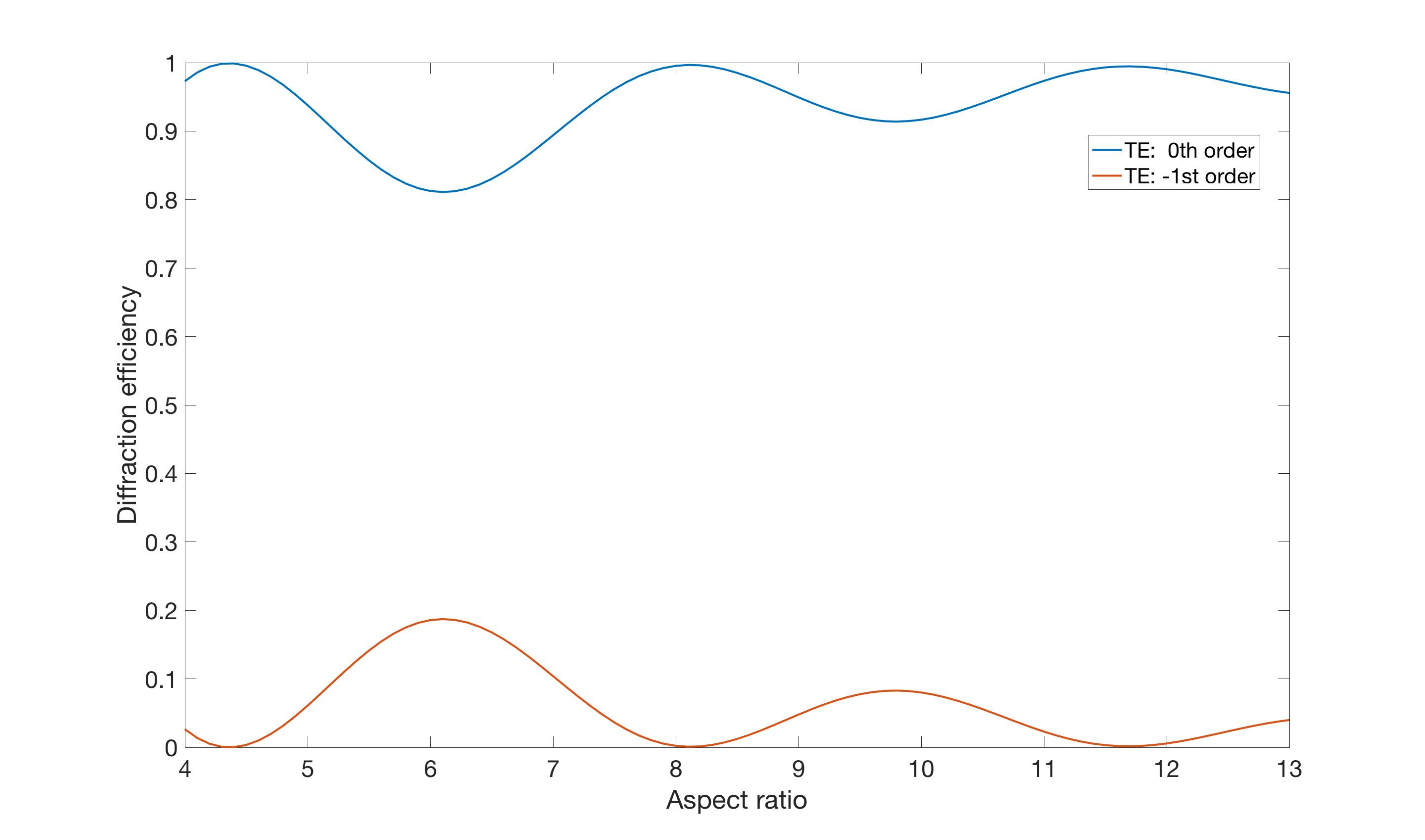}
\caption{Non-Littrow-mounting illumination condition}
\end{figure}

In this case, the maximum transmission ratio for the 0th diffraction beam is still high due to the low refractive index mismatch. However, the curves no longer follow the Mach-Zehnder interference relationship. The reason is that the 0th and -1st order diffraction beams are not symmetric in the non-Littrow-mounting condition. Another point of view to explain this phenomenon is to consider the grating as a volume hologram; whence, diffraction efficiency is influenced by the illumination angle. The Littrow-mounting illumination condition is also known as the Bragg-matched condition \cite{sun2015unified} where maximum diffraction efficiency is expected. For the non-Littrow-mounting case, the hologram becomes Bragg-mismatched and the diffraction will depend on the angular Bragg selectivity $\Delta \theta$, which is inversely proportional to the thickness of the volume hologram, i.e. $\Delta \theta \sim \frac{1}{H}$. Therefore, when the grating becomes thicker, $\Delta \theta$ is smaller and diffraction becomes more sensitive to illumination angle. In other words, when the grating becomes thicker,  the same non-Littrow-mounting illumination angle deviates more from the Bragg-matched condition resulting in even weaker diffraction.

\subsection {Truncated triangular case}
The structures that we analyzed so far are full duty-cycle triangular gratings with $f=0$ at the top and $f=1$ at the bottom, which have high transmission due to the low refractive index mismatch at both the input and output interface. Now, let us examine the cases where the structures are truncated triangular gratings [ $f\neq0$ at the top and/or $f\neq1$ at the bottom]. 

We still consider the Littrow-mounting illumination condition first. Here, we assume $f=0.3$ at the top and $f=0.7$ at the bottom. In this case, the structure is actually a trapezoidal grating. The other parameters are set as: $L=600nm, n_{1}=n_{g}=1,n_{2}=n_{b}=1.45, \lambda=1064nm, \varphi_{in}=\sin^{-1}\left(\frac{\lambda}{2n_{1}d}\right)=62.46^{\circ}$. The diffraction efficiency plot with respect to the aspect ratio is shown in Fig. 5. 

\begin{figure}[h]
\centering\includegraphics[width=10cm]{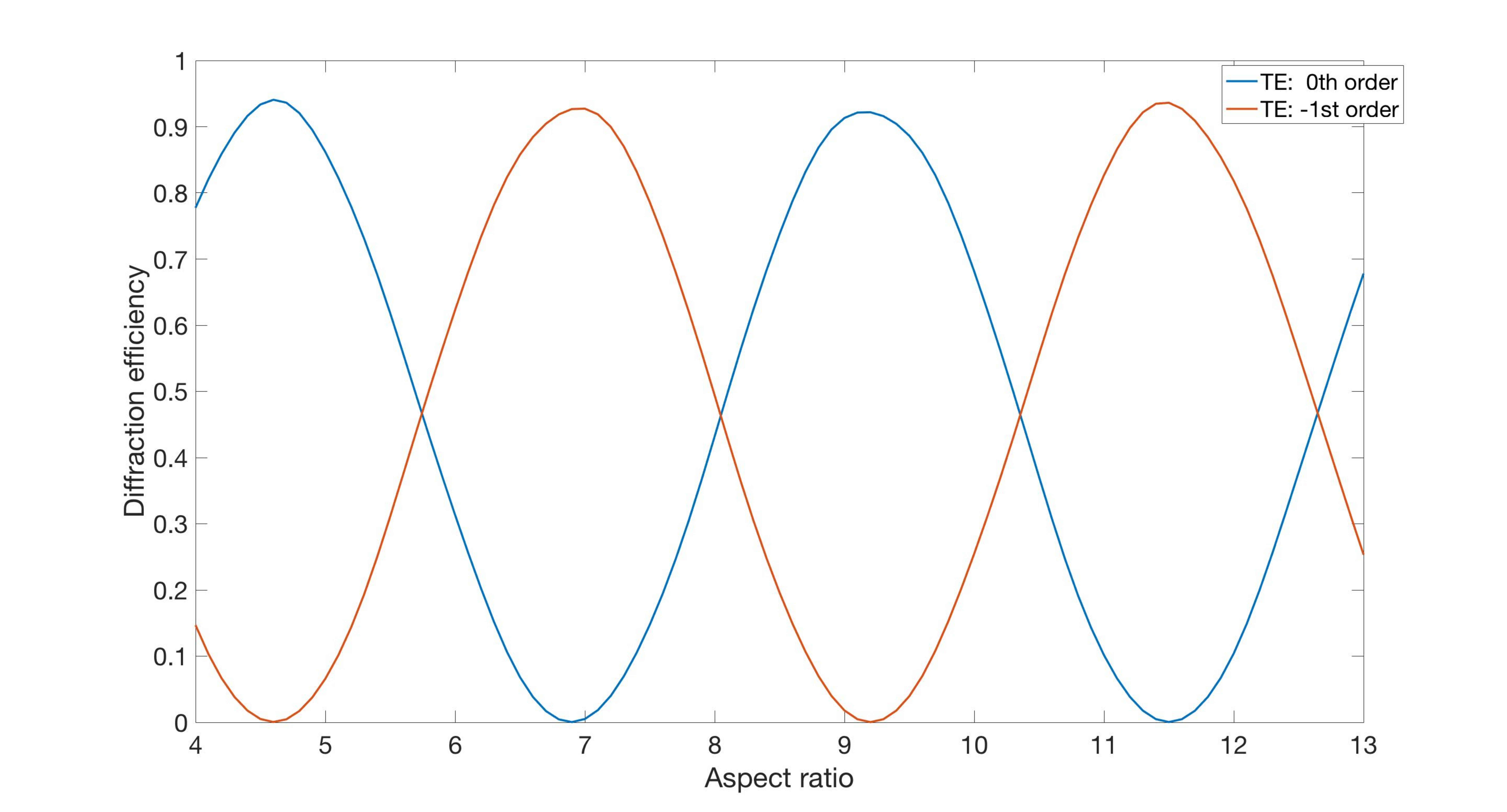}
\caption{Truncated triangular grating in Littrow-mounting case}
\end{figure}

From Fig. 5, we can observe  Mach-Zehnder type interference relationship: this is justified, since this is the Littrow-mounting illumination condition again. However, the transmission ratio is reduced compared to the full duty-cycle case [Fig. 3].  This result is also expected intuitively, as the reflection due to the refractive-index mismatch is higher in the truncated case. Moreover, this result actually highlights one advantage of our method over conventional SMM even in the Littrow-mounting case: when analyzing non-rectangular gratings in the Littrow-mounting case with conventional SMM, the diffraction efficiency is always computed in view of two-beam interference \cite{zheng2008polarizing}, where the reflections at the interfaces are not considered \cite{jing2013enhancement}. By contrast, TSMM inherently takes reflection into consideration and, therefore,  provides more accurate result.

We also analyze the truncated triangular grating under non-Littrow-mounting condition where $\varphi_{in}=\sin^{-1}\left(\frac{\lambda}{3n_{1}d}\right)$. We set the structural parameters as: $d=900nm, L=450nm, n_{1}=n_{g}=1,n_{2}=n_{b}=1.45$. The illumination wavelength is set to be $\lambda=1200nm$.  The diffraction efficiency plot with respect to the aspect ratio is shown in Fig. 6. 

\begin{figure}[h]
\centering\includegraphics[width=10cm]{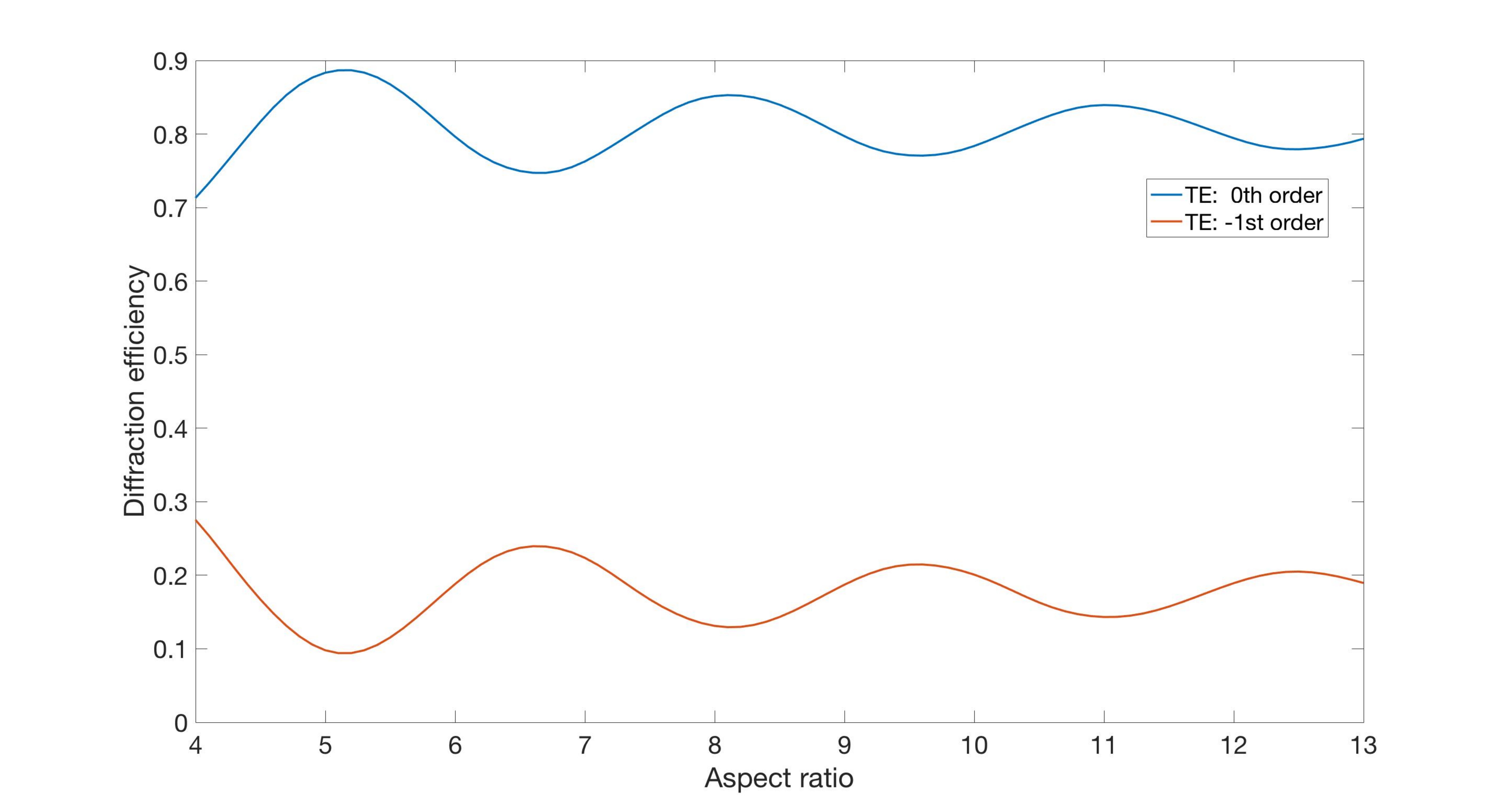}
\caption{Truncated triangular grating in non-Littrow-mounting case}
\end{figure}

As expected, the transmission of the 0th order beam is lower compared to the full duty-cycle case [Fig. 4] since the duty cycle at the bottom of the triangle is only $0.5$ now, which will cause a higher reflection at the output interface.

\subsection {Convergence of the method}
To demonstrate the stability of our method, we now examine the convergence of diffraction efficiencies with respect to the total number of layers $N$. Three different cases are considered: 

(1) Under Littrow-mounting illumination condition, the 0th order diffraction efficiencies of two gratings with different thickness are computed. The large aspect ratio grating has thickness $H=10\mu m$, while the small aspect ratio grating has thickness $H=6\mu m$. All other parameters are the same as those of the simulation for Fig .3.  

(2) Under non-Littrow-mounting illumination condition, the -1st order diffraction efficiencies of two gratings with different thicknesses are computed. The large aspect ratio grating has thickness $H=20\mu m$, while the small aspect ratio grating has thickness $H=14\mu m$. All other parameters are the same as those of the simulation for Fig .5. 

(3) Under non-Littrow-mounting illumination condition, the 0th order diffraction efficiencies of two gratings with different thickness are computed. The large aspect ratio grating has thickness $H=20\mu m$, while the small aspect ratio grating has thickness $H=14\mu m$. All other parameters are the same as those of the simulation for Fig .6. 

The convergence curves for all cases above are shown in Fig. 7. Here, all curves are normalized to their maximum respective values and then plotted in logarithmic scale. 

\begin{figure}[h]
\centering\includegraphics[width=12cm]{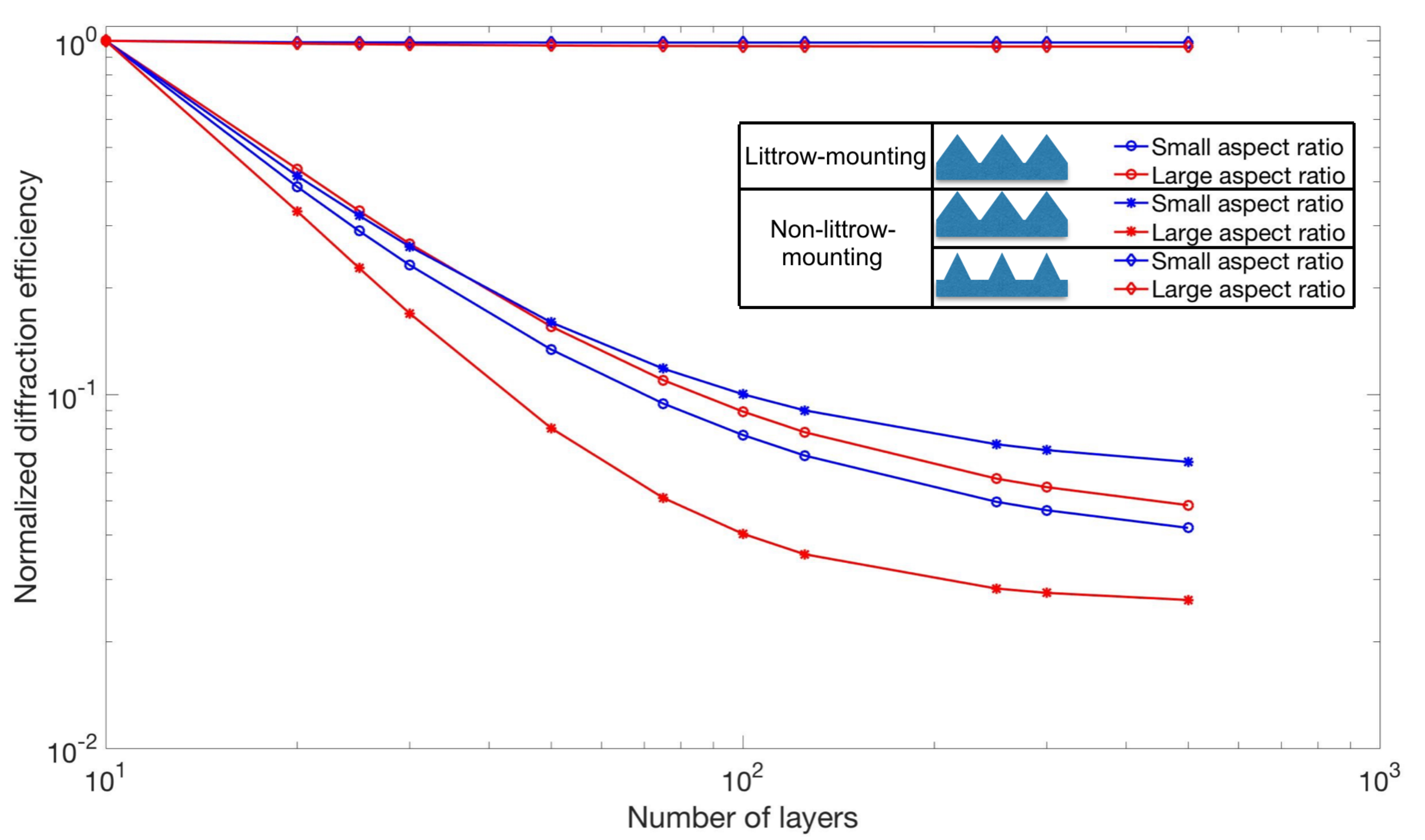}
\caption{Convergence curves}
\end{figure}

As expected, when $N$ increases ({\it i.e.}, the thickness of each layer decreases), the computed diffraction efficiencies in all the three cases converge, which demonstrates the stability of the proposed method. Also, we notice that we can not estimate the convergence rate merely from the aspect ratio of the grating. As we can see from Fig. 7, in case (1), large aspect ratio gratings have faster convergence rates, while in case (2) and (3), the small aspect ratio grating has a faster convergence rate. In fact, the convergence rate is influenced by the aspect ratio in two ways: on one hand, the larger aspect ratio means smaller slope, which is good for obtaining accurate result in TSMM, as we will explain in detail in the following section; on the other hand, larger aspect ratio means larger thickness of each discretized layer, which will in turn reduce accuracy.  Whether the slope or the layer thickness makes the dominant contribution to the convergence rate is determined by many other factors such as duty cycle and incidence angle.

\section{Slight taper approximation}
The validity of TSMM in computing the diffraction field of non-rectangular gratings has been demonstrated through a series of simulations. However, when applying this method in practice, we should be aware of the fact that the accuracy of this method is influenced by the taper angle $\Omega$ of the non-rectangular grating. An illustration of the taper angle $\Omega$ is shown in Fig. 8. This method works best when the slight taper condition is satisfied, i.e. $\Omega\ll1$. This requirement is actually due to three approximations that we make in developing TSMM.

\begin{figure}[h]
\centering\includegraphics[width=7cm]{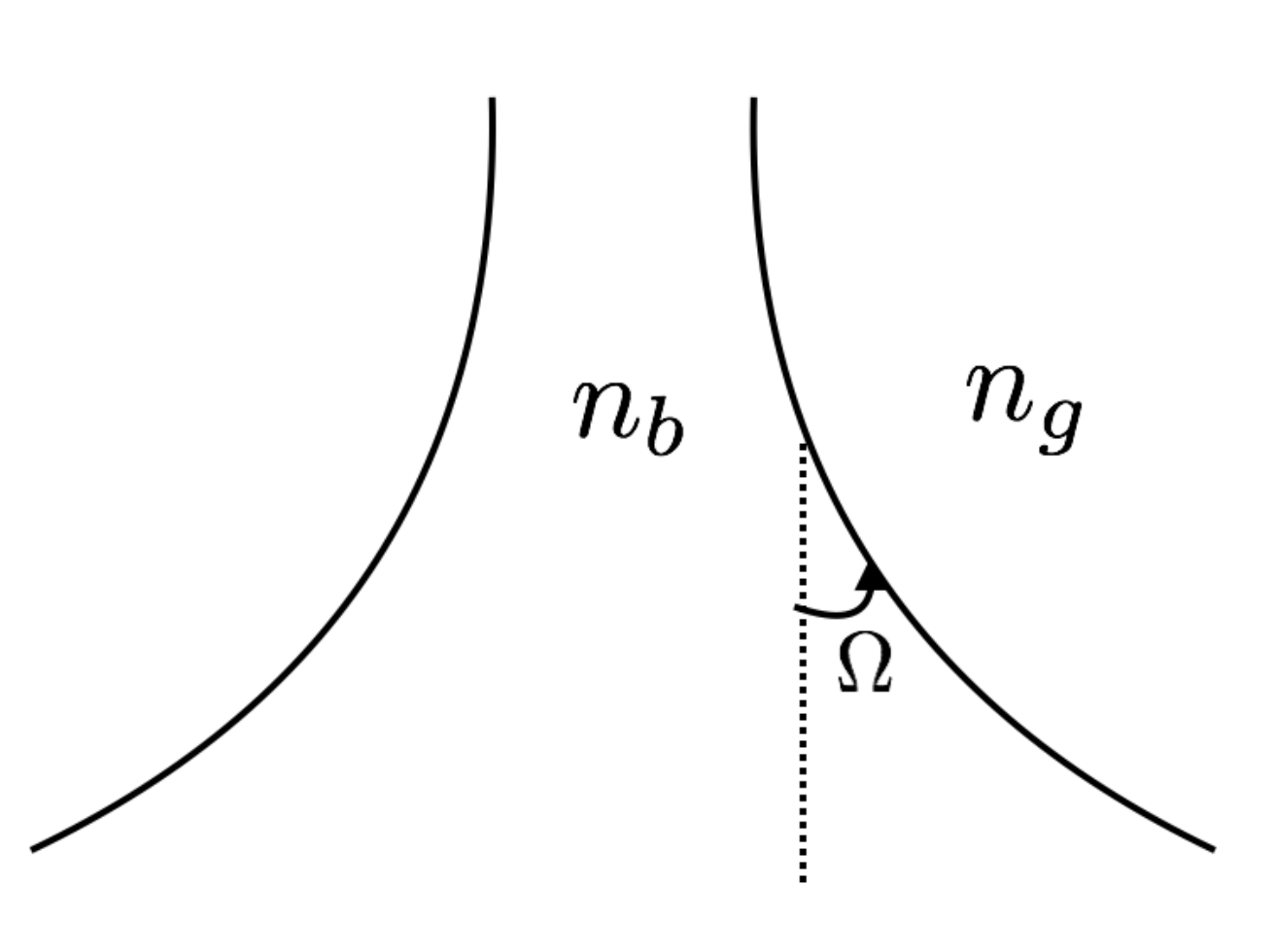}
\caption{Taper angle}
\end{figure}

(1) \textit{Propagating modes approximation}. In SMM, we consider the total field as the superposition of all propagating modes (with real $\neff$). By doing this, we actually eliminate all the evanescent modes (with imaginary $\neff$.) Thus, we introduce a truncation error. According to the adiabatic coupling theorem \cite{snyder1970coupling, johnson2002adiabatic}, coupling between propagating modes and evanescent modes is negligible when the taper angle is small enough. When the taper angle increases, the coupling also increases and the influence of evanescent modes becomes larger. For the triangular grating considered here, one quick way to estimate the truncation error is to compute the $1/e$ penetration depth $d_{p}$ for the first evanescent mode. The first evanescent mode has the smallest $|\neff|$ among all the evanescent modes, thus having the largest penetration depth. When $H\gg d_{p}$, we can consider coupling between propagating modes and evanescent modes to be very small and, therefore, justifiably negligible. In this way, we actually find a lower bound for the aspect ratio of the grating that our TSMM can provide accurate results. One way to improve the accuracy of our method is to take the evanescent modes into consideration \cite{yang2015evaluation}. We consider this improvement to be outside the scope of the present paper and defer to future work. 

(2) \textit{Uniform boundary condition approximation}. As can be seen from (4), we express the total fields $E_{t}$ inside the grating region as a superposition of grating modes $e_{q}$. Because of the non-rectangular structure, $E_{t}$ satisfies a non-uniform boundary condition at the ridge-groove boundary \cite{snyder1970coupling}. Nevertheless, the analytical functions of grating modes $e_{q}$, no matter if they are propagating or evanescent, are determined in the rectangular case, which satisfies the uniform boundary condition. Therefore, no finite summation of these grating modes could ever possibly satisfy the non-uniform boundary conditions. This fact, in turn, causes error in the computation of the non-TE case. Since the boundary conditions are satisfied to the order of $(n_{b}^{2}-n_{g}^{2})\cdot\tan\Omega\cdot E_{z}$, when $\Omega\ll1$ the corresponding error can be neglected, and then eq. (4) becomes valid.

(3) \textit{Partial derivatives approximation}. As shown in eq. (26), there are z-derivative terms inside the expressions of coupling coefficients. In TSMM, the grating is discretized along the z direction; this is how we can obtain $e_{q}(x,z_{i})$ and $h_{q}(x,z_{i})$ for each discretized layer $i$. To estimate the $z$-derivative terms, we fit those discretized values with polynomials to obtain analytical expressions $e_{q}(x,z)$ and $h_{q}(x,z)$ for each value of $x$ and then we compute the derivatives analytically. However, the accuracy of polynomial fitting is influenced by the taper angle. Intuitively, when the taper angle is large, the difference between two adjacent layers is large. As a result, the error in approximating the z-derivative terms also becomes large. To reduce this error, increasing the number of discretized layers is a possible approach. However, the penalty is the increase in the computational cost.

\section{Conclusion}
In this paper, we have demonstrated the Tapered Simplified Modal Method (TSMM) for grating analysis. To our knowledge, this is the first physically intuitive modal method reported that can be applied to non-rectangular gratings under arbitrary illumination condition, thus it greatly broadens the application of the modal method. The key element of TSMM is that we discretize the non-rectangular grating into several layers and treat each layer as a rectangular grating. Then, conventional SMM can be applied to each layer. With the help of tapered mode-coupling theory, we obtain the relationship between all the layers and, finally, we obtain diffraction and reflection coefficients for all orders. The validity of this method has been demonstrated through a series of numerical simulations. This novel method can help us better understand the processes that take place inside the grating region and facilitates the design of gratings with particular specifications.

\section*{Acknowledgments}
This work was supported by the U.S. Department of Homeland Security, Domestic Nuclear Detection Office, under the competitively awarded contract HSHQDC-13-C-B0040. This support does not constitute an express or implied endorsement on the part of the United States Government.

\end{document}